*Article*

# Innovative mini ultralight radioprobes to track Lagrangian turbulence fluctuations within warm clouds: electronic design

**Miryam E. Paredes Quintanilla[1,*], Shahbozbek Abdunabiev [1], Marco Allegretti [1], Andrea Merlone [2], Chiara Musacchio [2], Eros G. A. Pasero [1], Daniela Tordella [3], and Flavio Canavero [1]**

[1] Politecnico di Torino, Department of Electronics and Telecommunications (DET), corso Duca degli Abruzzi 24, 10129, Torino (Italy)
[2] Istituto Nazionale di Ricerca Metrologica, Str. delle Cacce, 91, 10135 Torino (Italy)
[3] Politecnico di Torino, Department of Applied Science and Technology (DISAT), corso Duca degli Abruzzi 24, 10129, Torino (Italy)

\* Correspondence: miryam.paredes@polito.it.

**Abstract:** Characterization of cloud properties remains a challenging task for weather forecasting and climate modelling as cloud properties depend on interdependent natural processes at micro and macro scales. Turbulence plays an important role in particle dynamics inside clouds; however, turbulence mechanisms are not yet fully understood partly due to the difficulty of measuring clouds at the smallest scales. To address these knowledge gaps, an experimental method for measuring the influence of fine-scale turbulence in cloud formation in-situ and producing an in-field cloud Lagrangian dataset is being developed by means of innovative ultra-light radioprobes. This paper presents the electronic system design along with the obtained results from laboratory and field experiments regarding these compact (diameter ≈ 30 cm), light-weight (≈ 20 g), and expendable devices designed to passively float and track small-scale turbulence fluctuations inside warm clouds. The fully customized mini radioprobe board (5 cm x 5 cm) embeds sensors to measure local fluctuations and transmit data to the ground in near real-time. The tests confirm that the newly developed probes perform well providing accurate information about atmospheric variables, referenced in space. The integration of multiple radioprobes allows for a systematic and accurate monitoring of atmospheric turbulence and its impact on cloud formation.

**Keywords:** atmospheric probe; instrumented balloon; research sonde; atmospheric turbulence; warm clouds; Lagrangian measurements; wireless sensor network; LoRa; low-power sensors.

## 1. Introduction

Clouds are a natural complex feature of Earth and a key element in climate change and climate sensitivity, since their characteristics directly influence the global radiation budget, the global hydrological cycle (through precipitation), and the atmospheric dynamics [1,2]. Clouds cover approximately two thirds of the globe at any time, and they are the principal source of uncertainty in future climate and weather projection [3–6]. This is because clouds involve processes on a vast range of spatial and temporal scales, ranging from the order of few microns, where droplets nucleate and collide-coalesce, to the thousand-of-kilometers extent of global circulation [6]. Clouds represent a substantial challenge for scientific understanding and modelling, since the available methods are yet not able to characterize the entire cloud system and related interactions across scales.

Both, the intense turbulence of the airflow hosting the clouds and the less intense turbulence that characterizes the environmental clear air surrounding them [7,8], play an important role in cloud evolution and related precipitations. Nonlinear dynamical processes of vortex stretching, entrainment and mixing greatly influence the nucleation of water droplets and associated evaporation-condensation and collision-coalescence [7]. To address these knowledge gaps, different laboratory experiments, field observations and numerical simulations have been undertaken, to understand cloud microphysics and, particularly, the inherent turbulence interactions. Investigation methods include remote sensing by means of radars and lidars [9,10], in-situ observations including manned and unmanned airborne platforms (airplanes, helicopters, tethered lifted systems, etc.) [11–13], laboratory experiments in wind tunnels and climate chambers [14–16], and numerical simulation experiments carried out via Navier-Stokes direct numerical simulation of small portion of clouds [17,18].

We present here an in-situ method for measuring the influence of fine-scale turbulence in cloud formation, which is based on the design and implementation of an innovative ultra-light (about 20 grams) biodegradable and expendable radiosonde here referred as radioprobe. A radiosonde is a battery-powered instrument carried into the atmosphere usually by a weather balloon with radio transmitting capabilities [19]. The idea was developed during the proposal writing of a European Horizon 2020 Marie Sklodowska Curie project which was approved in 2016 (H2020 MSCA ITN ETN COMPLETE, GA 675675: Innovative Training Network on Cloud-MicroPhysics-Turbulence-Telemetry [6]). The mini radioprobes are used to passively track turbulent fluctuations of air velocity, water vapor and droplets concentration, temperature and pressure in warm clouds and surrounding ambient air according to the Lagrangian description [20] of turbulent dispersion, as proposed by Richardson in 1926 [21,22].

These compact light-weighted devices with maximum target weight of 20 grams and diameter of 30 cm, are designed to float at altitudes between 1-2 km and be alive for approximately 1 hour. The radioprobes are capable of passively tracking small-scale turbulence fluctuations inside warm clouds and surrounding air since they can be considered as markers in a Lagrangian description of the airflow. In order to enable them to float, the radioprobe electronics are housed inside 30 cm diameter balloons made of biodegradable materials, which are filled with an adequate mixture of helium gas and ambient air to reach a buoyancy force equal to the system weight. Considering that the floating devices will not be recovered once they have finished their mission, the design accounts for the use of environmental-friendly materials to minimize any possible negative impact on the environment. To this end, the external balloons are made of biodegradable materials tailored to provide hydrophobicity and flexibility properties [23]. In the context of research balloons, these innovative devices can be catalogued as mini ultralight instrumented weather balloons. However, they are different from other instrumented devices developed for atmospheric sounding, like the NCAR-NOAA Global Hawk tethered dropsonde (weight 167 g, length 30.5 cm, diameter 4.6 cm; square-cone parachute: 20 cm on a side) used for vertical atmospheric profiling measurements (no Lagrangian trajectories) and launched by an unmanned aircraft from the National Aeronautics and Space Administration (NASA) [24], or the NOAA ground-launched smart balloon ( diameter of 335 cm) housing the sensors inside the enclosure and used for Lagrangian experimental hurricane research [25]. Additional devices are the short-range ground-launched weather balloon from NOAA carrying a tethered radiosonde (balloon diameter about 152 cm) [26], and the air-released balloon tethered microsonde (total weight 65.6 g) for supercell thunderstorm studies [27].

The Lagrangian balloons described in this article behave as instrumented particles embedding a set of compact size sensors for the measurement of local fluctuations of temperature, pressure, humidity, acceleration and trajectory. They can be released into the atmosphere from unmanned aerial vehicles or small airplanes. During the flight, the smart radioprobes acquire, pre-process, store, arrange and transmit in real time the obtained data to different ground stations located on earth through a dedicated long-range saving-power wireless radio transmission link [28].

This paper focuses entirely on the electronics design of the new radioprobe and is organized as follows. Section II describes the radioprobe environment and addresses the design requirements. Section III describes the system architecture and the design methodology. Section IV reports on the performance evaluation. Section V labels conclusions and presents future work.

**2. Understanding the sensor environment and design requirements**

The mini probes are conceived to work at the microphysical level and measure small-scale turbulence fluctuations inside warm clouds. To this end, they must have unique characteristics that allow them to behave as instrumented particles and track Lagrangian trajectories once being released into the atmosphere. This specific kind of radioprobe must be as small as possible to have a minimal inertia and a minimal size (diameter) compared to the expected trajectory length and be able to passively follow cloud fluctuations. To float on an isopycnic surface, the density of the radioprobe must correspond to the density of air at the target flight altitude (between 1 km and 2 km). To this end, the weight and volume of the radioprobe's balloon must remain relatively unaltered for the duration of the flight as presented in an initial study of the balloon materials in [23]. Based on that, the size required for the instrumented balloon to float was determined by the Archimedes' principle for buoyancy. The spherical balloon size must be about 30 cm in diameter. It should be noted that we foresee to operate our mini-green radioprobes in a spatial volume that has a limited extension, a few kilometers along the horizontal and maximum a few hundred meters along the vertical. This is a physical space that includes both the cloud, or part of it, and a part of the surrounding under-saturated air. The turbulence that characterizes this system is three-dimensional, and not necessarily includes large scale coherent vortices as those which are typical of the coherent vortices in rotating barotropic flows [29,30]. The light-small-green-expendable radioprobes of which we here describe the electronics and telecommunication project are used to study a few aspects associated to the microphysics of lukewarm clouds that is conditions typically far from those met in geophysical rotating turbulent flows. The lower limit of eddies size intended to observe in the atmosphere is in the range of 0.5 - 1 m, with a frequency around 0.5 – 1 Hz, and a kinetic energy per unit of mass in between 0.001 – 0.01 $(m/s)^2$. The higher limit is around a few kilometers, which brings about frequencies as low as $10^{-4}$ Hz.

Since a large number of radioprobes is required for the scope, they should be low cost. Although current radioprobe manufacturing and launch procedure (either from ground or aircraft) are relatively inexpensive [31], the miniaturization of these innovative devices, together with the non-necessity of a mother aircraft, expensive ground launch station or complex logistics, will further reduce costs associated to their production and release.

Each device must include different sensors to measure velocity, acceleration, vorticity, pressure, temperature and humidity fluctuations inside warm clouds. According to the environmental conditions that can be found inside real clouds, the operational requirements for the radioprobe sensors can be summarized as follows: external temperature: range from 0 °C to +30 °C, external relative humidity (RH): range from 0 % to 100 % RH, external pressure: range from 400 mbar to 1100 mbar, trajectory: +/-100 mm accuracy, and total wind fluctuation: max 20 m/s inside a cloud.

The data collected during the flight must be sent wirelessly to a data acquiring system on Earth whilst the device is alive. For this purpose, a transmission technology able to reach relatively long distances without consuming much power is required.

## 3. Radioprobe system architecture and design methodology

The working principle of the entire system is shown in Figure 1. This Wireless Sensor Network (WSN) is structured in three main parts: the bio-balloon wrapped radioprobe, which includes the solid-state sensors to measure the physical quantities of interest and which transmits the collected and pre-processed data to ground (#1); the base stations, which receive, store and pass this information to the processing machine (#2); and the processing machine, which is used for database management, filtering and visualization (#3).

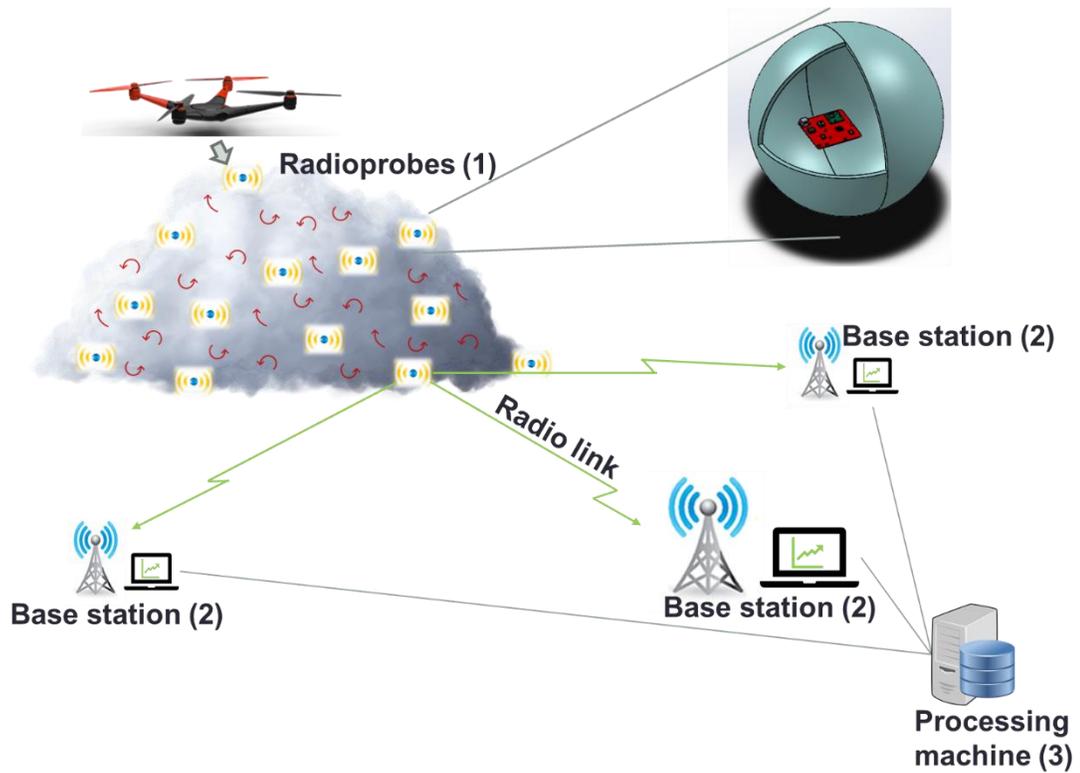

**Figure 1.** Working principle and radioprobe system architecture.

The block diagram of the radioprobe is illustrated in Fig. 2, where the system is represented by its several functional units: a data processing and control unit (1), a radiocommunication system (2), a temperature, pressure and humidity sensor stage (3), a positioning and tracking sensor stage (4), and a power supply unit (5).

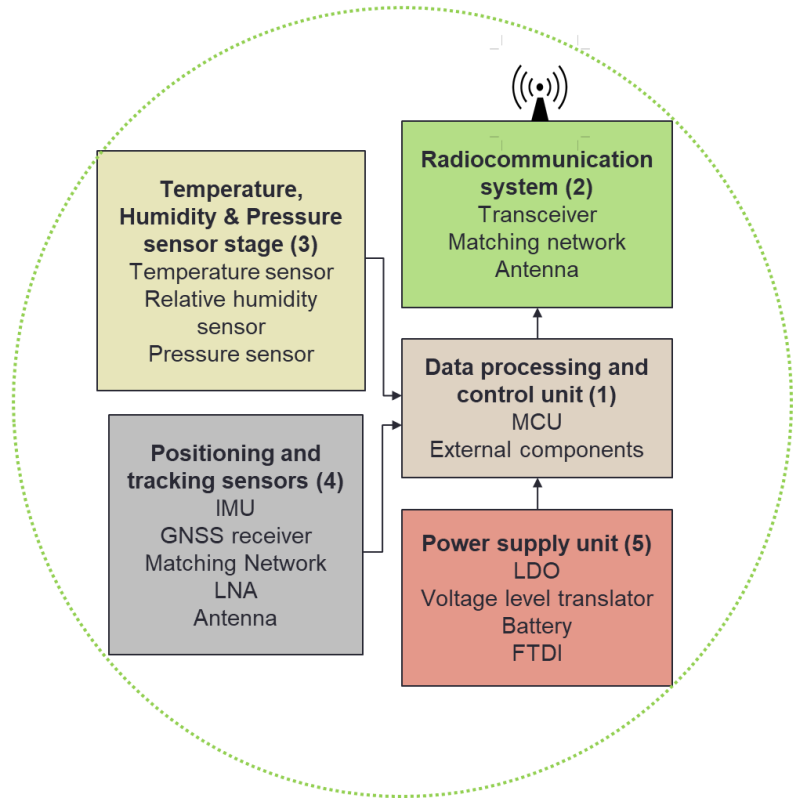

**Figure 2.** Block diagram of the mini radioprobe.

The printed circuit board (PCB) realization of the radioprobe is displayed in Fig. 3. All the electronics are assembled on both sides of a 2-layer FR4 substrate with surface mount technology (SMD) components. It is a 50 mm x 50 mm rectangular structure with a thickness of 0.8 mm and weight of 7 g (without battery). The following subsections provide further details of each functional block of the mini probe and the ground station.

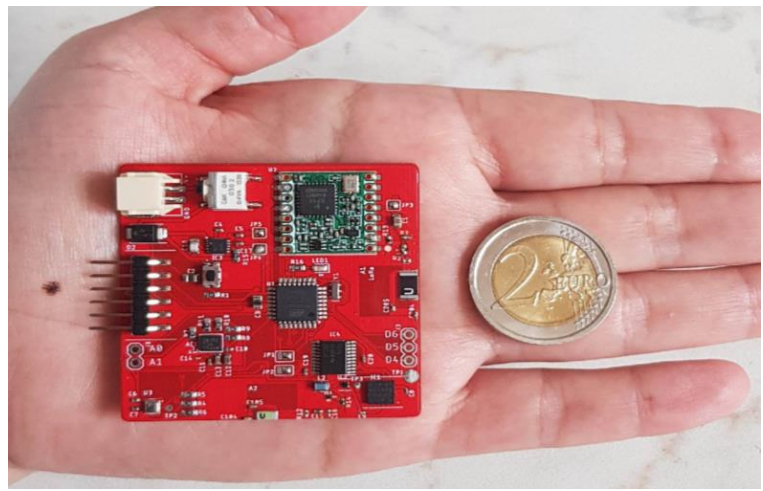

**Figure 3.** Top view of the PCB implementation of the radioprobe. Size 50 mm x 50 mm. Weight 7 grams.

*3.1. Data Processing and Control Unit*

The data processing and control unit block is the computational module of the radioprobe. It allows controlling and executing different subsystem processes in an automated way inside the device. In this unit, the data delivered by the sensors are interpreted, processed, saved and sent through the transmission module to the ground stations. For this purpose, the onboard low power complementary metal-oxide semiconductor (CMOS) 8-bit microcontroller ATmega328 from Microchip [32] has been selected as the central processing unit. It has 32 pins in a thin quad flat pack (TQFP) package with compact dimensions of 9 mm x 9 mm x 1 mm and weight of 70 mg. The microcontroller requires a supply voltage in the range from 1.8 V to 5.5 V and operates within temperature ranges from -40 °C to +85 °C. It requires low current consumption i.e., 0.2 mA in active mode, 0.1 µA in power-down mode and 0.75 µA in power-save mode @ 1 MHz, 1.8 V, 25 °C.

*3.2. Radio Communication System*

The radio communication system of the mini probes enables the one-way wireless communication with ground using radiofrequency signals. Due to the required criteria of the artificial floating probes, LoRa communication technology has been adopted. LoRa is a chirp spread spectrum (CSS) modulation technique, which encodes information in linearly increasing chirps [33,34]. LoRa was originally developed for the Internet of things (IoT) and since its release, it underwent enormous growth, being adapted for a wide range of applications [35]. Although LoRa is being used as part of the open-source LORAWAN specification, in this work it is used to create an ad-hoc private network and adapt the technology to the working scenario. To this end, the commercial off-the-shelf LoRa-based transceiver module RFM95 from HopeRF was used [36]. This transceiver and therefore the communication technology were previously tested by the authors under different scenarios [28,37–40]. It is a module featuring long-range spread spectrum communication links and high immunity to interference whilst optimizing the power use. This module allows power transmission ranges within 5 dBm (3.16 mW) to 20 dBm (100 mW), although according to the regulations released by the European Telecommunications Standards Institute (ETSI), the maximum power allowed in the European area is 14 dBm (25.12 mW) [41]. It requires a supply voltage in the range from 1.8 V to 3.7 V and operates within temperature ranges from -20 °C to +70 °C. The typical current consumption required by the transceiver are 0.2 µA in sleep mode, 1.5 µA in idle mode, 20 mA in transmit mode @ +7 dBm output power (OP), 29 mA in transmit mode @ +13 dBm OP, and 120 mA in transmit mode @ +20 dBm OP.

*3.3. Antennas*

Each tiny radioprobe includes two RF stages, one for the transmission of the in-flight collected data to ground, and one for the reception of positioning and timing data from satellites. The antennas used for the two stages are ceramic quarter wave chip antennas embedded in the system, one working in the LoRa sub-1GHz frequency band, and the other in the L1 frequency band, respectively. Both antennas used for the transmission and reception of the radioprobe data, are linearly polarized and have small dimensions, i.e., 5 mm x 3 mm x 0.5 mm, and 3.2 mm x 1.6 mm x 0.5 mm, respectively. They were mounted at the center of two different edges of the PCB top side and, since the chip itself is half of the antenna design, the bottom side of the PCB includes the ground plane layer to complete the antenna system. In addition, in order to minimize electric fields generated at the edge of the PCB and reduce crosstalk, via shielding was incorporated alongside the path of the RF signals and the ground clearance areas [42]. Moreover, with the purpose of ensuring the best possible RF performance, impedance matching practices were performed to ensure that most of the power is delivered between the transceivers and the antennas during the transmission and reception processes. The matching network extensions used for the antennas' tuning are L-section type, which uses reactive elements to match the load impedance to the transmission line.

*3.4. Temperature, Barometric Pressure and Relative Humidity Measurement*

After an extensive analysis of possible options and based on the physical constraints of the design, the combined module BME280 [43], which is a humidity sensor measuring ambient temperature, relative humidity and barometric pressure, was selected as the most suitable choice for the mini-probes. This all-in-one option consumes very little current (in the order of the µA), which makes it ideal for battery powered purposes as in the present case. The device comes in a land grid array (LGA) package of dimensions 2.5 mm x 2.5 mm x 0.93 mm, and requires a supply voltage in the range from 1.2 V to 3.6 V. The operating ranges of the device are 0 % to 100 % RH for relative humidity, 300 hPa to 1100 hPa for pressure, and -40 °C to +85 °C for temperature. In terms of overall performance, this device provides an maximum uncertainty of ±3 % RH and a resolution of 0.008 % RH for relative humidity, a maximum uncertainty of ±1 hPa and a resolution of 0.18 Pa for pressure, and a maximum uncertainty of ±1 °C and an output resolution of 0,01 °C for temperature measurements. The response time of the BME280 depends on the oversampling mode, selected filter and the data rate used. The oversampling modes available are 1, 2, 4, 8, and 16. The temperature, pressure and relative humidity measurements are extracted through reading commands implemented in the microcontroller. In the final radioprobe version, these sensors will be placed outside the balloon to be in direct contact with the atmosphere under study.

*3.5. Positioning and Tracking Measurement*

In the Lagrangian reference system, the fluid flow properties are determined by tracking the motion and properties of the individual fluid particles as they move in time [44]. For the radioprobe, the physical quantities already explained in the previous subsection, will be measured along the trajectory of the fluid particle as time passes. In this way, if many fluid particles (radioprobes) are tracked at the same time, the fluid properties for the whole domain can be obtained. The positioning and tracking electronic block allows to collect useful data to determine the trajectory and position followed by the radioprobe during its flight. The positioning and motion tracking is executed as a post processing task at the ground level and is obtained by sensor fusion algorithms based on Kalman and orientation filters. The orientation filter is used to fuse data coming from an inertial measurement unit IMU, and the Kalman filter exploits the output of the orientation filter and fuses it with the data coming from a Global Satellite Navigation System (GNSS) receiver.

The IMU used for this block is the nine-axis inertial module device LSM9DS1 [45] that combines a three-axis digital linear acceleration sensor, a three-axis digital angular rate sensor, and a three-axis digital magnetic sensor, all in a single package. It comes in a compact LGA package of dimensions 3.5 mm x 3 mm x 1.0 mm, requires a supply voltage in the range from 1.9 V to 3.6 V, and operates within temperature ranges from -40 °C to +85 °C. The device has a linear acceleration measurement range of ±2, ±4, ±8, ±16 g, a magnetic field full scale of ±4, ±8, ±12, ±16 gauss, and an angular rate full scale of ±245, ±500, ±2000 dps. The output data rate configuration modes available for the IMU sensors are: 10 – 952 Hz for the accelerometer, 14.9 – 952 Hz for the gyroscope, and 0.625 – 80 Hz for the magnetometer. The typical current consumption required by the IMU when operating in normal mode is 600 µA for the accelerometer and magnetic sensors, and 4 mA for the gyroscope @ 2.2 V, T = 25 °C. The main function of the IMU unit is to provide force, angular rate, orientation information of the radioprobe flight.

The GNSS receiver unit used in this block is a professional ultra-small, super low power System-in-Package (SiP) ZOE-M8B [46] module that offers a Super-Efficient (Super-E) mode option for improving the power consumption. It comes in an advanced soldered land grid array (S-LGA) package of dimensions 4.5 mm x 4.5 mm x 1.0 mm, requires a supply voltage in the range from 1.71 V to 1.89 V, operates within temperature ranges from -40 °C to +85 °C, and draws low current i.e., 34.5 mA for acquisition, 32.5 mA for tracking (continuous mode), 7.3 mA (Super-E mode), and 6.3 mA (Super-E mode power save) @ 1.8 V, 25 °C. For GPS and GLObal NAvigation Satellite System (GLONASS), the GNSS receiver provides a horizontal position accuracy of 3.5 m (Super E-mode), 2.5 m (continuous mode), and 4.0 m (Super E-mode power save), with a maximum navigation update rate of 10 Hz for continuous mode and 4 Hz for Super-E mode. This receiver module can measure dynamics up to 4 g, at altitudes up to 50 km and velocities up to 500 m/s. The GNSS is connected to the microcontroller through a bidirectional voltage-level translator, which serves as an interface for the different voltage requirements. The GNSS signal input is attached to an additional external low noise amplifier (LNA) for best performance in terms of noise figure and robustness against jamming, RF power and Electrostatic Discharge (ESD). The main function of the GNSS unit is to provide periodic reference position information of the radioprobe flight for removing drifts in the IMU output. Since the GNSS receiver consumes relatively higher power than the other sensors, the Super E-mode combined with periodic off and on periods of the GNSS module are used to save power.

*3.6 Power supply unit*

Power consumption is a critical key of the radioprobe design since it is closely related to the total weight of the device. The power supply block provides the electric power to the system and incorporates two different options to energize the circuit. The first option consists of a single non-rechargeable battery used to provide enough power to the electronic circuit while keeping the whole system light and autonomous during the flight. To this purpose, a single 4.0 V Lithium Metal Oxide (LMO) battery with nominal capacity of 125 mAh and pulse current capacity of 3.75 A is used. The cell's weight is 9 g with a volume of 3.2 cm3 and wide temperature operating range of -55 °C to +85 °C. This battery complies with Underwriters Laboratories Inc. (UL) safety standards. It is made of non-toxic and non-pressurized solvents and includes less reactive materials than standard lithium cells. The second option includes a FTDI USB to serial connection used mostly for code uploading and management purposes. To provide the required supply voltages (3.3 V and 1.8 V) to the different components, the circuit incorporates the dual low-dropout (LDO) voltage regulator LP3996SD [47], which can source 150 mA and 300 mA at an ultra-low quiescent current of 35 μA.

**4. Experimental results and discussion**

This section reports on the outcomes of the different experiments performed to validate the radioprobe system. The performance of the system was assessed based on communication reliability, sensor reliability, and power consumption.

*4.1 Antenna Matching and Data Transmission Ranges*

To improve the radioprobe antenna system performance, the antennas' characterization was done by measuring their complex impedance values and adjusting the matching network components to obtain an acceptable S11. To this end, the portable USB Vector Network Analyzer (VNA) Keysight P9371A, was employed. Since the antenna impedances were not matched to 50 ohms as expected, the L-type matching components were calculated based on the normalized load impedance and then soldered on the PCB to improve the quality of the match. Moreover, the resonance frequency of the antennas was shifted to the desired ones (around 868 MHz and 1575 MHz). The results of the matching and frequency tuning procedures for both, the transmission and reception RF stages, are shown in Table 1.

**Table 1.** Results of the matching and frequency tuning procedures

| Frequency [MHz] | Initial $S_{11}$ [dB] | Final $S_{11}$ [dB] |
|---|---|---|
| 865.2 | -0.59 | -23.99 |
| 868.0 | -0.56 | -21.09 |
| 1575.0 | -1.22 | -23.09 |
| 1602.0 | -1.22 | -17.34 |

As a result of this process, the performance of both antenna systems was considerably improved. The initial reflection coefficients of the system were enhanced by approximately 40 times for the transmission RF stage and 19 times for the receiving RF stage thus, ensuring in this way the maximum power transfer in the RF units.

In addition, with the goal of testing the communication system of the radioprobe, some sets of measurements using different network configurations were carried out. The initial field measurement (Setup 1, Figure 4) included propagation measurements using a point-to-point static network configuration in an urban environment to identify the transmission ranges of the system in harsh propagation conditions. This test was carried out in the city of Turin - Italy, specifically within our University and its surroundings. The network setup included a radioprobe (transmitter) creating and sending a unique sensor identification (ID) together with a counter, and a ground station (receiver) receiving and storing the messages. The aim of the counter was to identify the losses of packets having a known progressive number included in the data frame. The transmitter was located at eight different positions from P1 to P8, while the receiver was located at a fixed position Rx. Also, at the receiver side, a Spectrum Analyzer (SA) model R&S ZVL was placed to measure the power of the signal spectrum; however, for most of the points, the noise floor of the instrument was higher than the incoming signal thus the measurement of the power spectrum was not possible. This behavior emphasizes the robustness of LoRa technology and the opportunity to establish communication links in challenging environments. The receiver module was programmed in order to provide useful information about the signal quality, that is, signal-to-noise ratio (SNR) and received signal strength indicator (RSSI) of the packets. The receiver was placed at an approximated height of 17 m and the transmitter at a height of 1 m above the street level. The tests were made using a programmed output power of 10 dBm, central frequency 865.2MHz, spreading factor of 10, and a bandwidth of 125kHz. The set of analyzed data consisted of blocks of 200 packets for each transmitter position. The fixed location of the ground station and the different positions of the transmitter (radioprobe) are shown in Figure 4. The obtained results of the measurements are reported in Table 2.

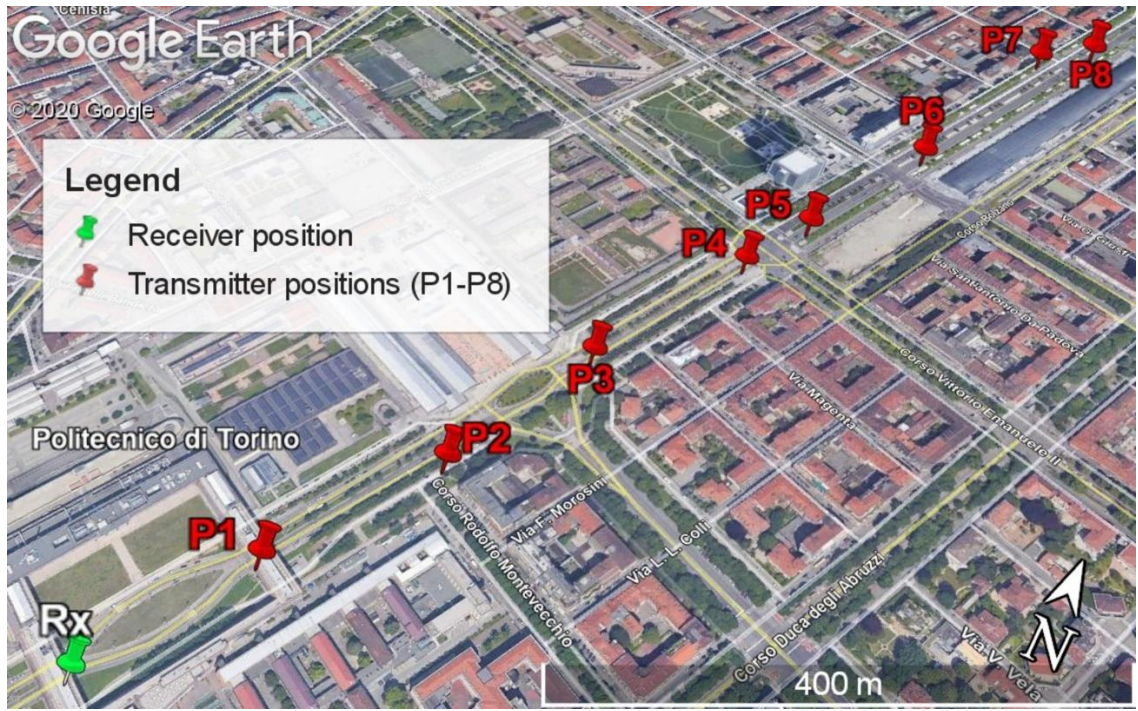

**Figure 4.** System setup 1 used to determine the transmission ranges reached by the radioprobe system in an urban environment, displayed on a map. Transmitters (P1 to P8) and receiver (Rx) position, with relative distance indications. Google earth view.

**Table 2.** Results of the point-to-point measurements in urban environment (Setup1).

| Tx Position | Distance [m] | SNR mean [dB] | RSSI mean [dBm] | Received packets [%] |
|---|---|---|---|---|
| P1 | 138 | 7 | -95 | 100.0 |
| P2 | 280 | 2 | -113 | 99.5 |
| P3 | 455 | -7 | -123 | 99.5 |
| P4 | 648 | -9 | -124 | 77.8 |
| P5 | 737 | -2 | -120 | 99.5 |
| P6 | 905 | -9 | -125 | 96.0 |
| P7 | 1173 | -13 | -122 | 95.5 |
| P8 | 1232 | -12 | -124 | 52.0 |

As result of these propagation measurements, different transmission links were tested to understand the transmission ranges that can be reached by the system, of course, in a more difficult environment where partial or total obstruction of the Fresnel zone is present. The closest eight different transmitter positions (P1 to P8) were selected since the percentage of received packets was greater than 50 %. The maximum propagation distance tested was 1232 m of distance between the transmitter and the receiver. In most positions, the communication link was affected by direct obstacles and reflections from diverse sources, which is a common propagation issue in built-up areas. For all the measurements, the SNR ranged from +7 dB at the nearest distances to -13 dB at the longest ones. The negative SNR values obtained is an inherent LoRa characteristic, which indicates the ability of this technology to receive signal power below the receiver noise floor [48]. As expected, the RSSI of the packets decreased with distance and non-line-of-sight (NLOS) between the transmitter and the receiver; however, for most of the cases, the percentage of received packets was higher than 95 %. These measurements provided a good reference of possible transmission ranges that can be achieved by the radioprobes when floating into the unobstructed free atmosphere environment.

A second field measurement included propagation measurements using a point-to-point dynamic network configuration in an open area environment (Setup 2, Figure 5). Unlike the previous experiment, the mini radioprobe transmitting the information was attached to a reference radiosonde, which was part of an automatic atmospheric sounding system, to simulate similar conditions in which the radioprobes will be released. This experiment was carried out at the Cuneo-Levaldigi meteorological station (id LIMZ) of the Regional Agency for the Protection of the Environment (ARPA) of Piedmont, Italy, where an atmospheric balloon is launched into the atmosphere twice a day. The sounding system consisted of a large Helium-filled balloon of about 1.5 m of diameter, tethering through a polypropylene string a Vaisala RS41 radiosonde able to provide temperature, humidity, wind, height and pressure information through a telemetry link to ground.

The network setup for this measurement included a fully operational mini radioprobe gathering, processing, packing and transmitting the information from the different sensors, and a ground station receiving, storing and post-processing the received messages. The tiny radioprobe was attached to the front side of the reference radiosonde's cover and activated just before the launch to save energy for the flight. The radioprobe's transceiver was programmed to provide an output power of 14 dBm at a central frequency of 865.2 MHz, spreading factor of 10, and bandwidth of 125kHz. The receiver was placed close to the ground at an approximated height of 1 m. Since this set of measurements were carried out in a non-obstructed open environment, the transmitter was in LOS with the receiver at all positions. The system setup and trajectory followed by the systems with respect to the ground station are shown in Figures 5 and 6 respectively.

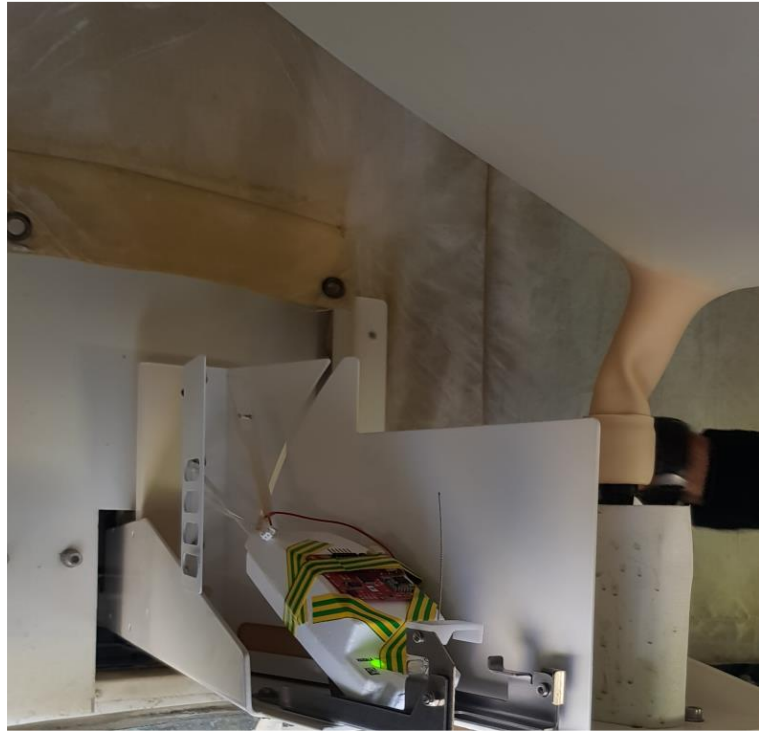

**Figure 5.** System setup 2 used to determine the transmission ranges reached by the radioprobe system in an open area environment. Tiny radioprobe attached to the reference atmospheric sounding system.

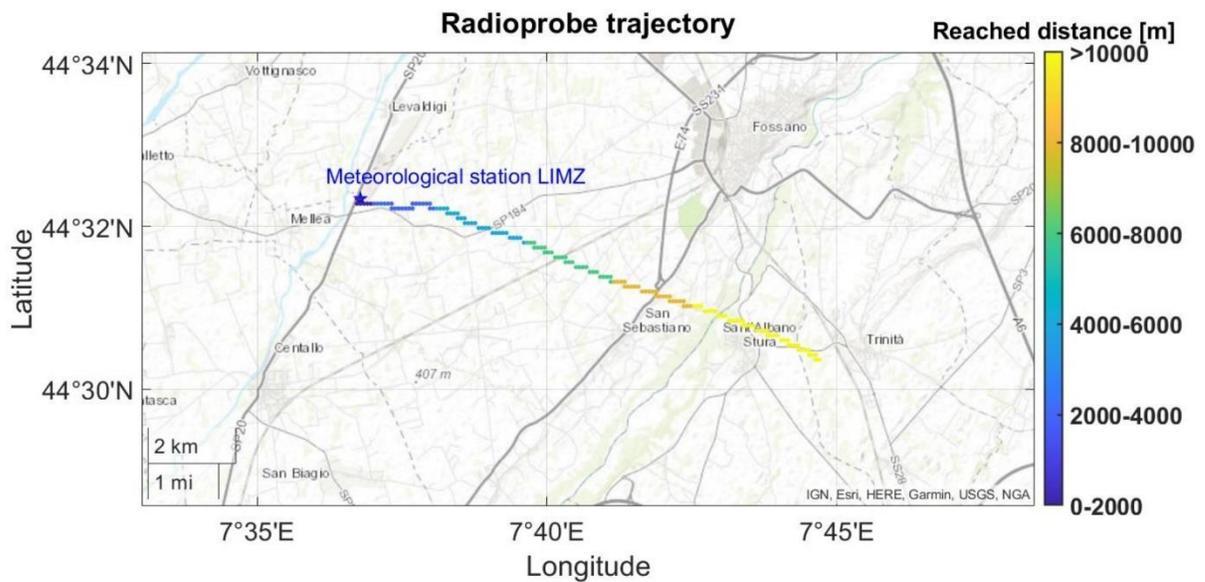

**Figure 6.** Trajectory of the fully operational radioprobe attached to the reference atmospheric sounding system, displayed on a map. The color bar indicates the separation distance reached by the system with respect to the ground station.

As result of these propagation measurements, the maximum transmission range reached by the radioprobe system in an open environment was determined. Although the reference atmospheric sounding system was intended for vertical atmospheric profiling measurements of the troposphere and low stratosphere, and not for warm cloud environments with heights between 1 km and 2 km, it provided a good reference to test our system in a dynamic atmosphere environment free of obstacles. A summary of the obtained results of the measurements is reported in Table 3.

Table 3. Results of the point-to-point measurements in open environment (Setup2).

| Distance [m] | SNR mean [dB] | RSSI mean [dBm] | Total transmitted packets | Number of received packets | Received packets [%] |
|---|---|---|---|---|---|
| Up to 1000 | 5 | -95 | 40 | 37 | 92.5 |
| Up to 2000 | 4 | -99 | 103 | 98 | 95.2 |
| Up to 3000 | 2 | -102 | 156 | 146 | 93.6 |
| Up to 4000 | 2 | -103 | 210 | 196 | 93.3 |
| Up to 5000 | 1 | -104 | 243 | 226 | 93.0 |
| Up to 6000 | 1 | -104 | 276 | 240 | 87.0 |
| Up to 7000 | 0 | -105 | 297 | 259 | 87.2 |
| Up to 8000 | 0 | -105 | 322 | 283 | 87.9 |
| Up to 9000 | -1 | -106 | 348 | 294 | 84.5 |
| Up to 10000 | -1 | -106 | 376 | 296 | 78.7 |
| Up to 11000 | -1 | -106 | 449 | 297 | 66.2 |
| Up to 14000 | -1 | -106 | 462 | 298 | 64.5 |

The reference atmospheric sounding system carrying the attached mini radioprobe reached a maximum height of approximately 32 km and a horizontal range of approximately 108 km before the balloon burst. The tiny radioprobe reached a maximum height of approximately 11 km, a horizontal range of 7 km and a straight distance of 13 km before losing contact with the ground station. In total, 462 packets were sent from the mini radioprobe during the flying time for a time span of approximately 22 minutes after the launch.

For all the measurements, the SNR ranged from +5 dB at the nearest distances to -1 dB at the longest ones. As expected, the RSSI of the packets decreased with the increase in distance between the transmitter and the receiver. Although there was an intermittency in the reception of some packets due to the high ascending velocity of the sounding system, the percentage of received packets for the first 5 km was higher than 90 %. This is a good indicator for a warm-cloud monitoring system where the intended observation heights are between 1 and 2 km with much lower fluctuation velocities.

The communication technology was also used to demonstrate that the materials used for the bio envelope of the radioprobe is sufficiently transparent to radio waves and does not hamper the electromagnetic transmission; however, this study will be fully described in a future paper related to the biodegradable balloon development.

*4.2 Sensors testing and validation*

For the purpose of properly calibrating and validating the temperature and humidity sensors' response, a set of tests were carried out in the Applied Thermodynamics Laboratory of the Italian National Metrology Institute (INRiM). A climatic chamber Kambic KK190 CHLT specifically developed for meteorology and climate metrology was used [49]. It allows temperature regulation in the range from -40 °C to 180 °C, and relative humidity control in the range from 10 % to 98 % RH. The reference temperature values were obtained through four platinum resistance thermometers (Pt100) calibrated in INRIM laboratory placed inside the climatic chamber, Pt100 are read using external precision Super-Thermometer FLUKE 1594a. The reference humidity value was obtained with a Delta Ohm humidity and temperature probe calibrated at INRIM connected to a datalogger model HD27.17TS. The uncertainty of the Pt100 ranges from 0.011 °C for positive temperatures and 0.020 °C for negative temperatures. The total uncertainty of the Delta Ohm probe declared is ±3 % RH.

In order to test not only the accuracy of the temperature and humidity radioprobe sensors but also to have an idea of the possible spread of their behavior, three radioprobe electronic boards were used for this experiment.

They were placed inside the climatic chamber, together with reference temperature sensors and humidity probes for comparison purposes. The temperature and relative humidity measurements from the BME280 were extracted through reading commands implemented in the microcontroller through the I²C communication interface at a sampling frequency of 1 Hz.

The climate chamber was set at temperature of +20 °C and a relative humidity of 30 % RH as initial configuration. Then, additional controlled variations of chamber environment in terms of temperature and humidity were applied. In the first test small incremental steps in temperature of 2 °C were realized (keeping RH at 30 %) until reaching T = +24 °C, each one for a time span of approximately 30 minutes. After that the climatic chamber was configured to provide larger controlled variations in temperature starting from the current set values T = +24 °C, RH = 30 % until reaching -5 °C, 0 °C and 10 °C. Temperature steps need a time span of approximately 1 hour each to obtain temperature stability of the whole system. This temperature cycle was done in order to simulate conditions faced by the radioprobes on site. Although warm clouds are composed only of liquid water having temperatures above 0 °C (32 °F), the cycle also included negative temperature values to test the sensors' performance under extreme situations. The measurement results obtained in the second test are shown in Figure 7.

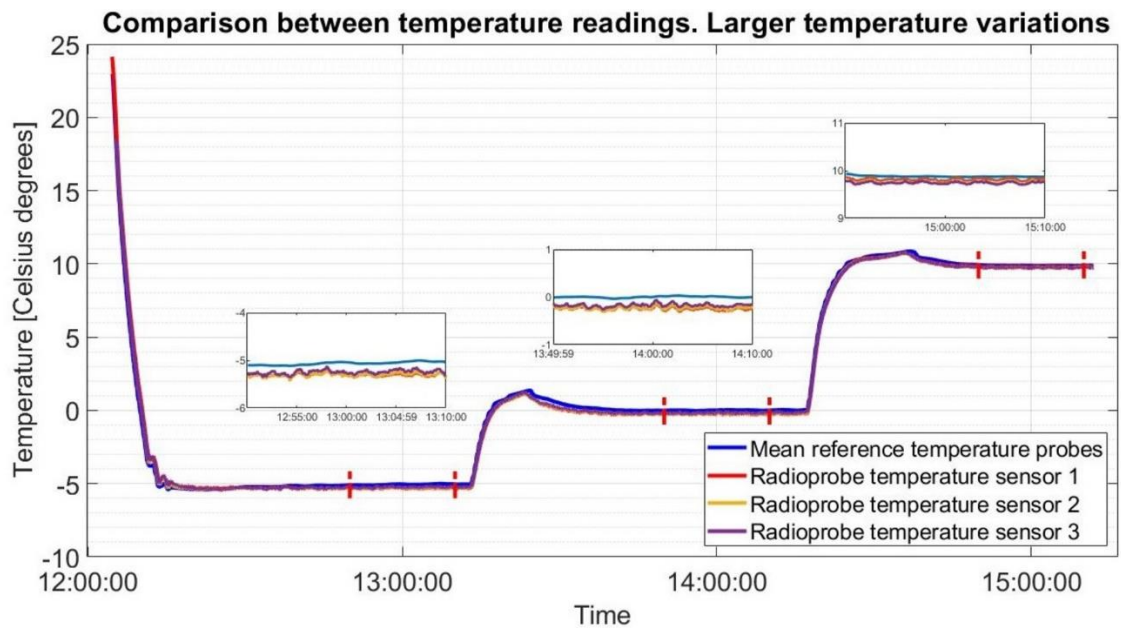

**Figure 7.** Comparison of temperature measurements between reference temperature sensors and radioprobe sensors. Climatic chamber is set to provide controlled variations in temperature starting from T = +24 °C, RH = 30 % until reaching the set points of T = -5 °C, T = 0 °C and T = 10 °C.

In the third test the relative humidity was changed from 10 % RH to 20 %, 40 % and 60 %, at a constant temperature of +30 °C; each step needs a time span of approximately 30 minutes. In order to statistically compare the obtained data, the Makima interpolation technique, which is an algorithm for one-dimensional interpolation, was used considering, at each set point, approximately 5 minutes of data selected when temperature and humidity conditions inside the chamber are stable. The statistical results of the second and third tests are shown in Table 4 and Table 5.

**Table 4.** Statistical comparison between radioprobe sensors and INRIM reference sensors readings. Temperature

| | | Radioprobe 1 | Radioprobe 2 | Radioprobe 3 |
|---|---|---|---|---|

| Temperature set point for test [°C] | Temperature measured by reference sensors (mean) [°C] | Mean [°C] | Mean error[1] [°C] | Standard deviation[2] [°C] | Mean [°C] | Mean error[1] [°C] | Standard deviation[2] [°C] | Mean [°C] | Mean error[1] [°C] | Standard deviation[2] [°C] |
|---|---|---|---|---|---|---|---|---|---|---|
| -5 | -5.063 | -5.31 | 0.25 | 0.04 | -5.30 | 0.24 | 0.04 | -5.25 | 0.18 | 0.04 |
| 0 | 0.002 | -0.25 | 0.25 | 0.03 | -0.23 | 0.23 | 0.03 | -0.17 | 0.18 | 0.03 |
| 10 | 9.878 | 9.82 | 0.065 | 0.02 | 9.75 | 0.13 | 0.03 | 9.74 | 0.13 | 0.02 |

[1] Temperature difference between reference sensor reading and the radioprobe sensor reading
[2] Standard deviation of radioprobe temperature reading

**Table 5.** Statistical comparison between radioprobe sensors and INRiM reference sensors readings. Relative Humidity

| RH set point for test [%RH] | RH measured by reference sensors (mean) [%RH] | Radioprobe 1 | | | Radioprobe 2 | | | Radioprobe 3 | | |
|---|---|---|---|---|---|---|---|---|---|---|
| | | Mean [%RH] | Mean error[1] [%RH] | Standard deviation[2] [%RH] | Mean [%RH] | Mean error[1] [%RH] | Standard deviation[2] [%RH] | Mean [%RH] | Mean error[1] [%RH] | Standard deviation[2] [%RH] |
| 10 | 10.50 | 13.12 | 2.62 | 0.01 | 14.74 | 4.24 | 0.02 | 14.16 | 3.66 | 0.02 |
| 20 | 19.75 | 19.85 | 0.09 | 0.08 | 21.35 | 1.60 | 0.17 | 21.09 | 1.34 | 0.18 |
| 40 | 37.68 | 35.31 | 2.37 | 0.10 | 35.64 | 2.04 | 0.12 | 36.06 | 1.62 | 0.12 |
| 60 | 59.70 | 56.13 | 3.57 | 0.07 | 54.53 | 5.17 | 0.05 | 55.69 | 4.01 | 0.04 |

[1] Relative humidity difference between reference sensor reading and the radioprobe sensor reading
[2] Standard deviation of radioprobe relative humidity reading

As result of this experiment using a high-precision climatic chamber and calibrated reference sensors, the performance of the radioprobe sensors was evaluated. The behavior of the radioprobe sensors lies between the specifications given by the manufacturer for most of the cases (i.e. temperature accuracy ±1 °C, relative humidity ±3 % RH). There are a few exceptions for the relative humidity measurements that might be caused by the uncertainties introduced by the reference sensor itself (accuracy of the humidity reference sensor ±3 % RH).

An additional field experiment was carried out to verify the response of the temperature, pressure and humidity sensor stage nested within the radioprobe board. The data obtained came from the experiment setup using the ARPA sounding system already described in the subsection 4.1: Antenna Matching and Data Transmission Ranges. The fully operational mini radioprobe was fixed to the front side of the reference Vaisala RS41-SG radiosonde case with the help of a non-conductive adhesive tape. It was constantly measuring, processing, packing and transmitting the information to the base station located on ground. The reference probe incorporated a temperature sensor using a linear resistive platinum technology, a humidity sensor integrating humidity and additional temperature sensing elements, and a GPS receiver allowing the derivation of pressure, height and wind data [50]. Regarding the accuracy provided by the reference instrument, the uncertainties declared for sounding are 0.3 °C for temperature measurements (for sounding lower than 16 km), 4 % RH for humidity measurements, and 1.0 hPa / 0.5 hPa for pressure measurements (for pressure values greater that 100 hPa). In order to statistically compare the obtained data, the set of measurements considered for the analysis corresponds to the interval up to which the percentage of received packets was greater than 90 %. At this point, the straight distance between the flying system and the base station was approximately 5 km. The measurement results obtained are shown in Figure 8. The statistical results of this test are shown in Table 6.

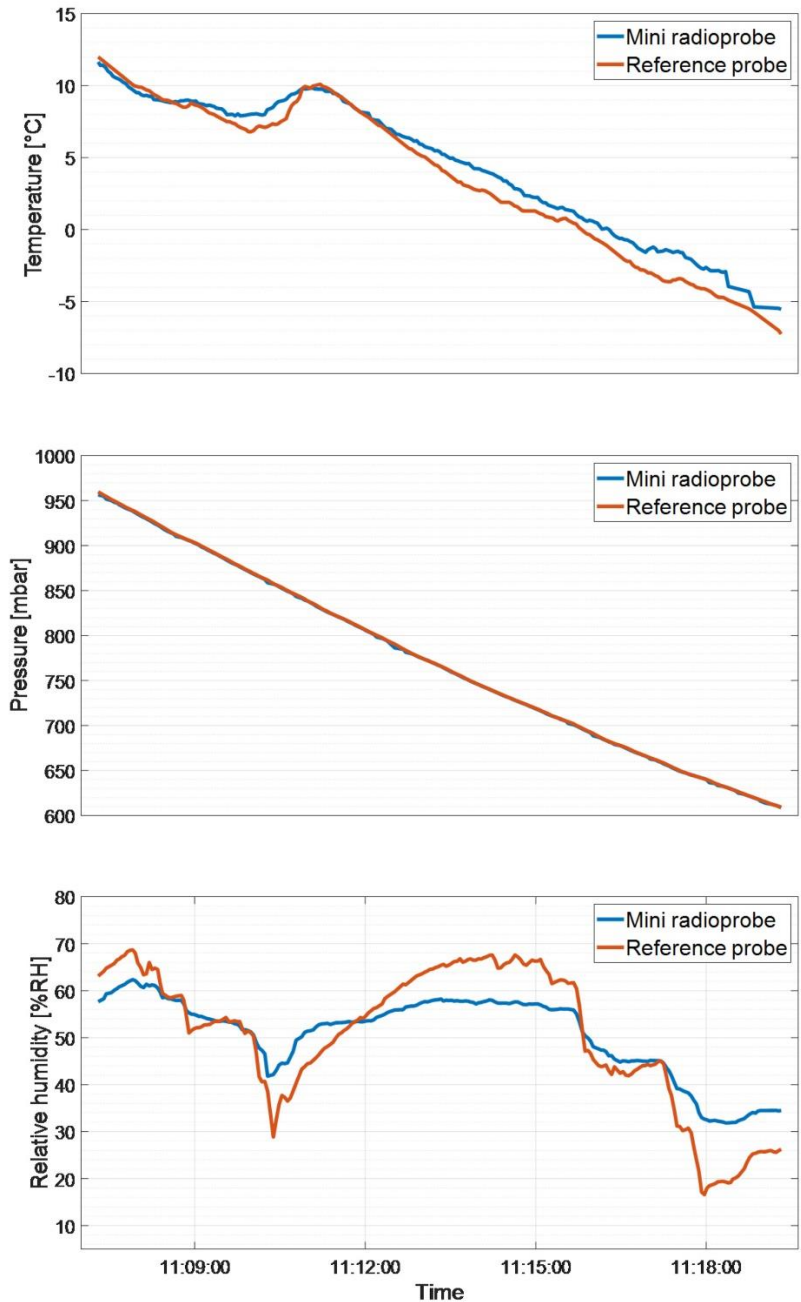

**Figure 8.** Comparison of temperature, pressure and relative humidity measurements between the ARPA reference sonde sensors and the radioprobe sensors.

**Table 6.** Statistical comparison between radioprobe sensor readings and ARPA reference sonde readings.

|  | VAISALA reference sensor measurements (mean) | Radioprobe measurements | | |
|---|---|---|---|---|
|  |  | Mean | Mean error[1] | Standard deviation[2] |
| **Temperature [°C]** | 4.16 | 4.93 | 0.87 | 0.56 |
| **Pressure [mbar]** | 774.14 | 773.53 | 0.63 | 0.58 |
| **Relative humidity [%RH]** | 50.74 | 50.86 | 5.53 | 3.71 |

[1] Difference between reference sensor reading and the radioprobe sensor reading
[2] Standard deviation of radioprobe temperature reading

As result of this experiment using a VAISALA radiosonde as a reference, the performance of the radioprobe's temperature, humidity and pressure sensor block was evaluated. From Figure 8 it is possible to observe some differences between the measurements provided by the radioprobe sensors and the reference instrument. These effects could have been produced by the position itself of the mini radioprobe onto the case containing the reference probe. Due to the lack of enough space available for placing the radioprobe and to avoid its fall during the flight, it was tightly attached to the reference probe leading to potential undesired effects. For instance, being in direct contact with the main body of reference instrument case, certainly the energy dissipated by the reference probe could have affected the radioprobe measurements. Also, since the air-flow in direction to the vent-hole of the TPH sensors was partially obstructed, the exchange of sufficient air was not possible contributing to errors in the measurements. Notwithstanding the aforementioned and, considering the limited resources in the design (e.g. small size, ultra-light weight, low power and low-cost sensors), it can be said from the obtained results that the performance of the TPH radioprobe sensors is good enough for the purpose of the radioprobe development. Overall, and considering the uncertainties introduced by the reference sensors, the behavior of the TPH radioprobe sensors lies between the specifications given by the manufacturer as can be seen in Table 6.

Future experiments will include a different setup of the instruments to overcome the problems encountered during the execution of this field experiment.

For the purpose of validating the positioning and tracking radioprobe sensors unit, a field experiment using as reference the GPS positioning data logs from a smartphone device was performed. This test was carried out in an open-area of the city of Turin. The system setup included a radioprobe measuring and partially processing the readings from the IMU sensors (accelerometer, gyroscope and magnetometer), and gathering the geolocation and time updates from the GNSS receiver. The radioprobe was configured in order to provide a GNSS sensor update each two seconds and two IMU sensor updates every second. It was connected via serial port to a portable PC for the data logging. Additionally, an Android-based smartphone model Samsung Galaxy S8+ executing a GNSS logger application for recording the position and path followed was used. This application provided positioning updates for each second.

For this experiment, the radioprobe and the smartphone simultaneously recorded data during a walk. Before starting the measurements, the calibration of the IMU sensors was performed to ensure that the readings and the output of the pre-filtering process executed at the radioprobe side are accurate. While being at rest, the bias errors and noise introduced by the accelerometer, gyroscope, and magnetometer were properly identified. In addition, since the GNSS update frequencies between the reference and radioprobe were different, the IMU readings were used to predict positioning information for the intermediate time steps. To this end, the IMU sensor data were processed using Madgwick filtering, which is an orientation algorithm to describe orientation in three-dimensions [51], to get accelerations in NED (North, East, and Down) absolute frame. This frame is useful for the post processing analysis to predict the radioprobe's position along its trajectory. Thus, acceleration data in absolute frame can be combined with LLA (Latitude, Longitude, and Altitude) absolute positioning data coming from the GNSS server. In this way, it is possible to have 5 (1 GNSS update and 4 predictions with IMU data) positioning information for each 2 seconds. The raw acceleration data along x, y and z directions in the radioprobe's body frame, and the converted acceleration in absolute frame after applying the orientation filter are shown in Figure 9. Since the experiment was performed in a horizontal plane, it is possible to see the north and east accelerations around zero, except for small fluctuations due to walk maneuver. Instead, for the down direction, the acceleration was around 10 m/s$^2$ because of gravity.

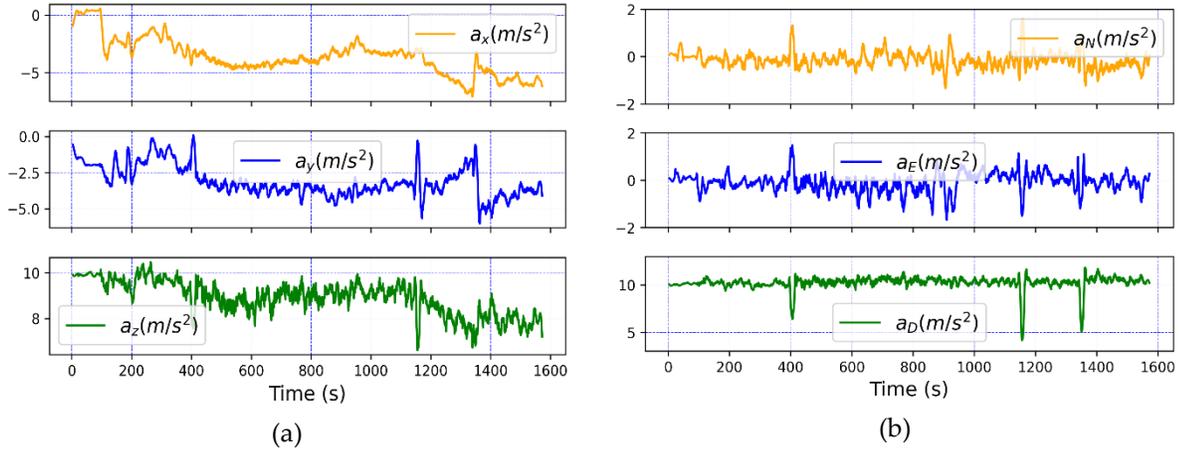

**Figure 9.** Absolute acceleration: (**a**) Raw acceleration in radioprobe's body frame; (**b**) Filtered acceleration in NED frame.

During the experiment, the total travelled distance from the starting to the final points was approximately 1.6 km for a time span of approximately 30 min. The trajectory recorded by both systems together with the comparison between trajectories along north (Latitude) and East (Longitude) directions are shown in Figure 10. The statistical results of the positioning sensors accuracy (IMU and GNSS) are shown in Table 7.

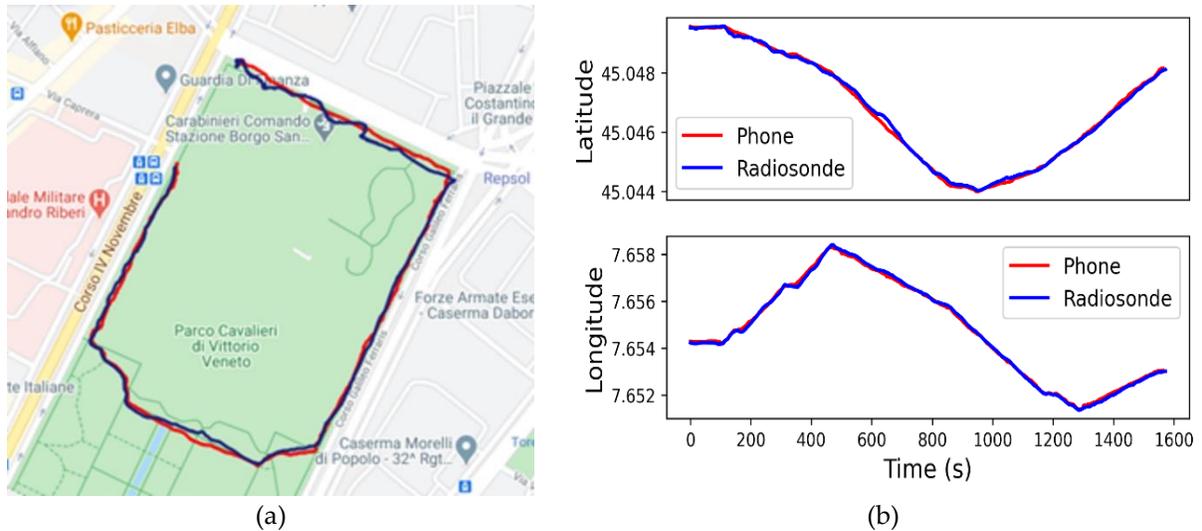

**Figure 10.** Trajectory recorded by the radioprobe (blue line) and the smartphone GPS logger (red line) during a walk: (**a**) Trajectory comparison displayed on a map; (**b**) Latitude and longitude comparison between the radioprobe and the smartphone GPS logger in function of time.

**Table 7.** Statistical results of positioning sensor accuracy (IMU and GNSS) during the experiment

| Sensor | | Offset (sensor bias) | Standard deviation |
|---|---|---|---|
| Accelerometer [m/s$^2$] | x | 0.26 | |
| | y | 0.21 | 0.025 |
| | z | -0.45 | |
| Gyroscope [degree/s] | x | 1.03 | |
| | y | 1.22 | 0.1 |
| | z | 8.80 | |
| | x | 84.56 | 4.2 |
| Magnetometer | y | -211.68 |  |
| [mGauss] | z | -271.32 |  |
| GNSS | Latitude | - 8.80E-06 | 5.73E-05 |
| [degrees] | Longitude | -7.78E-06 | 7.40E-05 |

From the obtained results, it is possible to verify the reasonable performance of the positioning and tracking radioprobe sensor unit considering the limited resources at the radioprobe side (e.g. low power, low memory availability, light weight and not-expensive sensors). To overcome these challenges, the reduction of the IMU sampling rate and the activation of a GNSS super-saving mode (E-mode) are among the strategies used. The partially processed data generated at this stage constitutes the input for the further post processing step executed at the ground level to reconstruct the trajectory followed by the mini radioprobes.

An additional experiment to validate the positioning and tracking radioprobe sensor unit was conducted. Although the balloon's performance analysis is not the purpose of this work, we carried out a preliminary tethered balloon test at low altitude (30 - 50 m) to expose the radioprobe to real atmospheric air fluctuation and verify the fluctuation detection ability of the tiny radioprobe when flying. This test was carried out at Parco Piemonte, which is a wide tree-free park located at the south area of Turin. The field measurement consisted of a point-to-point dynamic network configuration including a fully operational radioprobe collecting and transmitting the about-flight information, and a ground station receiving, storing and post-processing the received messages. The mini radioprobe was inserted in the middle of the Helium-filled biodegradable balloon and released into the low atmosphere. In order not to lose the measuring system, the balloon was attached to a long thin thread and held by one of the participants. The radioprobe's transceiver was programmed to provide an output power of 14 dBm at a central frequency of 865.2 MHz, spreading factor of 10, and bandwidth of 125kHz. The receiver was placed close to the ground at an approximated height of 1 m and at an approximate distance of 25 m from the initial balloon release point. Both the transmitter and the receiver were in LOS during the execution of the experiment. The trajectory followed by the radioprobe during the flight is shown in Figure 11.

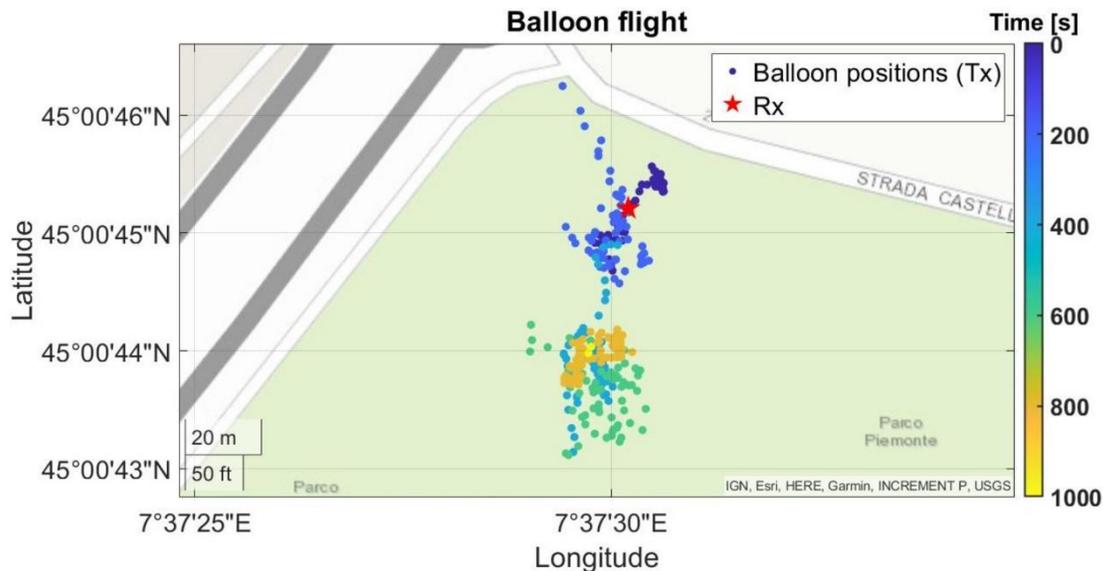

**Figure 11.** Low-atmosphere trajectory of the fully operational radioprobe inserted in a Helium-filled biodegradable balloon, displayed on a map. The color bar indicates the elapsed time.

The IMU measurements (acceleration, angular rate and magnetic field) are displayed in Figure 12.

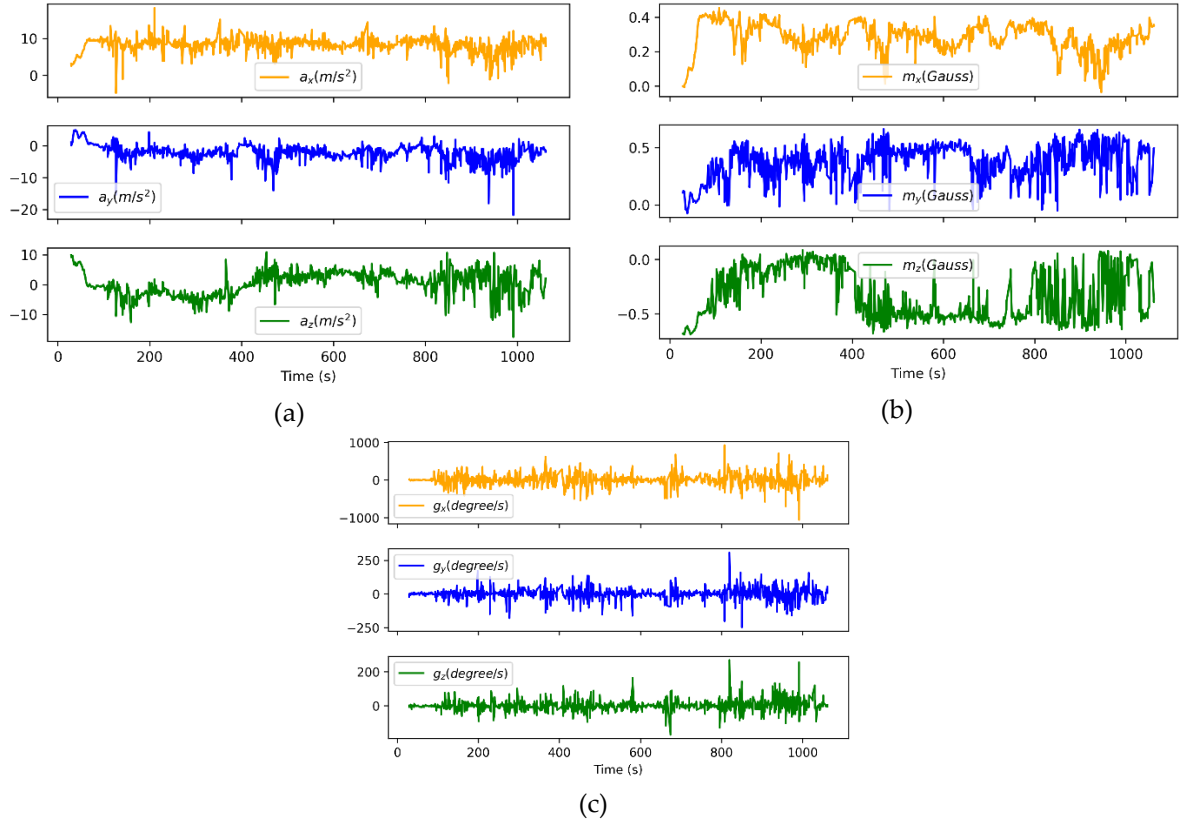

**Figure 12.** Radioprobe sensor measurements sent to the ground station for post-processing purposes: (**a**) Raw acceleration measured by the radioprobe sensors during the flight; (**b**) Raw magnetic field measured by the radioprobe sensors during the flight; (**c**) Raw angular rate measured by the radioprobe sensors during the flight.

As result of this experiment, the fully operational radioprobe was tested in a low-atmosphere open environment. The obtained results show the good radioprobe capacity to detect acceleration, angular rate and magnetic field fluctuations while flying inside the balloon in a dynamic environment. In addition, all the transmitted packets sent by the moving instrument were correctly received at the ground station. The SNR values ranged from +9 dB to -12 dB, and the RSSI of the packets from -65 dBm to -109 dBm.

*4.3 Power Consumption Analysis*

Power consumption is a key factor determining the radioprobe's life. In order to save energy, different solutions were adopted to extend the battery lifetime, according to the following power management strategies:

- Every single electronic component populating the radioprobe PCB was selected considering its power requirements to minimize the total energy consumption of the system.
- The readings obtained from the TPH and positioning/tracking sensor stages were partially processed at the radioprobe side to reduce the amount of information to be transmitted to ground.
- The number of packets to be transmitted were minimized by packing together two or more messages in a single data frame. In this way, the time-on-air of a single packet is higher, however, the number of transmissions is lower, hence saving power.
- Since the GNSS is the most power consuming sensor, it was periodically switched on and off to provide only the necessary information to update the reference position of the last Kalman's Filter output at the ground station level.
- The GNSS receiver was configured to work in the Super E-mode, which provides a good trade-off between current consumption and performance. This mode allows saving 3 times power compared with a traditional u-blox GNSS operating in full power mode [46] and, in addition, the receiver can automatically duty-cycle the external LNA to further reduce power usage.

As a result, the total current consumption of the radioprobe, which depends on the task in execution and the programmed transmission power, was properly measured. It may vary from an average value of approximately 90 mA to a maximum value of 123 mA when all the system is operating: the GNSS receiver is in acquisition mode, the radioprobe is transmitting a packet, the microcontroller is executing instructions, and the remaining sensors are taking measurements. In external conditions, the battery can supply energy to the radioprobe for approximately 60 minutes.

**5. Conclusions and future work**

This paper presents a novel method based on a WSN system for in-situ measuring the influence of fine-scale turbulence in cloud formation by means of the design and implementation of an innovative ultra-light expendable radioprobe. The integration of different areas of research for instance, low-power wireless sensor network communications, sensors and instrumentation for atmospheric measurements, sensors and instrumentation for trajectory tracking, antenna embedding and matching, and electronic board design, allowed the development of a complete and reliable system able to measure and transfer in an effective way atmospheric-based data through a long-range saving-power telemetry link to ground.

Outcomes from the different field measurements confirmed that the newly developed radioprobe device performs well and provides accurate information while keeping unique features for an instrumented weather balloon such as compact size, ultra-light weight, low-cost and low energy consumption. Each tiny probe can communicate correctly up to 5 km of distance, which is a transmission range enough for a warm cloud environment of heights between 1 and 2 km. With reference to the turbulence spectrum found in field measurements [52–56] and, considering the type of instrumentation embedded and the size associated to the radioprobes, it can be said that these devices can measure wavelengths in the order of 1 meter up to few kilometers, velocities from 30 – 50 cm/s up to 5 - 6 m/s, and accelerations up to ±4 g. In fact, the solid-state sensor producer datasheets [45,46] state that the IMU is capable of detecting linear accelerations up to ±4 g, and the GNSS receiver can work up to 4 g, at altitudes up to 50 km and velocities up to 500 m/s with the current configuration set in both devices. These findings suggest that these tiny radioprobes when embedded in a biodegradable balloon of diameter of 30 cm can behave as Lagrangian tracers of also small-scale turbulent fluctuations once released into warm clouds.

Based on the findings of the present paper, future work includes further miniaturization and weight optimization of the first radioprobe version here presented. In addition, the new electronic design will include a daughter board populated with the TPH sensors, which will be placed outside the enclosure to be in direct contact with the atmosphere and measure the physical parameters of interest. Furthermore, since the final goal of this research project is the generation of an in-field cloud Lagrangian dataset, the upcoming experiments will include a bunch of complete radioprobes (electronics and enclosure) working as a single system and transmitting simultaneously the collected cloud data to the ground stations for final post-processing tasks. Finally, for the purpose of recovering completely the power spectrum of the physical quantities under study inside clouds (temperature, pressure, humidity and acceleration), future experiments could include the use of more performing batteries (i.e. military grade), which will require the overcoming of acquisition administrative procedures.

**Author Contributions:** Conceptualization, M.E.P.Q., E.G.A.P., D.T. and F.C.; methodology, M.E.P.Q. and M.A.; software, M.E.P.Q. and S.A.; validation, M.E.P.Q, S.A., A.M. and C.M.; formal analysis, M.E.P.Q., S.A. A.M. and C.M.; investigation, M.E.P.Q. and M.A.; experimental design, M.E.P.Q., A.M., C.M. and D.T.; resources, M.A., A.M., C.M. and E.G.A.P.; data curation, M.E.P.Q.; writing—original draft preparation, M.E.P.Q.; writing—review and editing, E.G.A.P., D.T. and F.C.; visualization, M.E.P.Q. and S.A.; supervision, E.G.A.P., D.T. and F.C.; project administration, D.T.; funding acquisition, D.T. All authors have read and agreed to the published version of the manuscript.

**Funding:** "This project has received funding from the Marie - Sklodowska Curie Actions (MSCA) under the European Union's Horizon 2020 research and innovation programme (grant agreement n°675675)".

**Acknowledgments:** The authors would like to thank the Istituto Nazionale di Ricerca Metrologica (INRiM) for supporting the experimental measurements held at the Applied Thermodynamics Laboratory. The authors would like to thank the Regional Agency for the Protection of the Environment (ARPA) of the Piedmont Region of Italy for supporting the experimental measurements using their national atmospheric sounding system. The authors would like to thank the Istituto Italiano di Tecnologia (IIT) for supporting the development of the biodegradable balloons at the Smart Materials Department.

The authors would like to thank Professor Emilio Giovanni Perona and Dr. Silvano Bertoldo for providing useful suggestions during the conception and evolution of this research work. The authors would like to thank Dr. Athanassia Athanassiou, Dr. Giovanni Perotto and Eng. Giovanni Cipri for designing, characterizing and developing the probe envelope green material.

**Conflicts of Interest:** "The authors declare no conflict of interest."